\documentclass[%
reprint,
showpacs,%preprintnumbers,
amsmath,amssymb,
aps,
prl,
]{revtex4-1}
\usepackage{graphicx}% Include figure files
\usepackage{dcolumn}% Align table columns on decimal point
\usepackage{bm}% bold math
\begin{document}
\preprint{APS/123-QED} %or 'reprint' (mimics journal style); \texttt{letter} sized paper should
%be used when submitting to APS journals.
\title{Gene Expression Noise Facilitates Adaptation and Drug Resistance Independently of Mutation}
%\thanks{A footnote to the article title}%
\author{Daniel A. Charlebois$^{1,2}$}
\email{daniel.charlebois@uottawa.ca}
\author{Nezar Abdennur$^{2,3}$}
% \altaffiliation[Also at ]{Physics Department, XYZ University.}%Lines break automatically or can be forced with \\
\author{Mads Kaern$^{1,2,3}$}%
\email{mkaern@uottawa.ca}
\affiliation{%
$^{1}$Department of Physics, University of Ottawa, 150 Louis
Pasteur, Ottawa, Ontario, K1N 6N5, Canada.\\
$^{2}$Ottawa Institute of Systems Biology, University of Ottawa, 451
Smyth Road, Ottawa, Ontario, K1H 8M5, Canada.\\
$^{3}$Department of Cellular and Molecular Medicine, University of
Ottawa, 451 Smyth Road, Ottawa, Ontario, K1H 8M5, Canada.\\
% This line break forced with \textbackslash\textbackslash
}%
%\collaboration{MUSO Collaboration}%\noaffiliation
%\author{Charlie Author}
% \homepage{http://www.Second.institution.edu/~Charlie.Author}
%\affiliation{
% Second institution and/or address\\
% This line break forced% with \\
%}%
%\affiliation{
% Third institution, the second for Charlie Author
%}%
%\author{Delta Author}
%\affiliation{%
% Authors' institution and/or address\\
% This line break forced with \textbackslash\textbackslash
%}%
%\collaboration{CLEO Collaboration}%\noaffiliation
\date{\today}% It is always \today, today,
% but any date may be explicitly specified

\begin{abstract}
We show that the effect of stress on the reproductive fitness of noisy cell populations can be modelled as first-passage
time problem, and demonstrate that even relatively short-lived fluctuations in gene expression
can ensure long-term survival of a drug-resistant population. We examine how this effect contributes 
to the development of drug-resistant cancer cells, and demonstrate that permanent immunity can arise independently
of mutations.

%\begin{description}
%\item[Usage]
%Secondary publications and information retrieval purposes.
%\item[PACS numbers] 87.10.Ca,87.10.Mn,87.10.Rt,87.16.Yc,87.23.Cc, 
%87.23.Kg
%May be entered using the \verb+\pacs{#1}+ command.
%\item[Structure]
%You may use the \texttt{description} environment to structure your abstract;
%use the optional argument of the \verb+\item+ command to give the category of each item.
%\end{description}
\end{abstract}
%\pacs{...}% PACS, the Physics and Astronomy
% Classification Scheme.
%\keywords{Suggested keywords}%Use showkeys class option if keyword
%display desired
\maketitle
%\tableofcontents
Gene expression is a stochastic process that enables genetically identical cells in the same environment to exhibit phenotypic
variation~\cite{Kaern,Elowitz,Jia10}. This noise-induced non-genetic (epigenetic) variability can be beneficial to cell
populations experiencing acute stress by providing a temporary basis for natural selection~\cite{Blake,Fraser,Zhang,Zhuravel}. 

Experimental observations suggest that gene expression is inherently associated with `epigenetic memory', defined by the fluctuation relaxation time of a gene product within a cell lineage. In human lung cancer cells, this relaxation time can be as long as four generations~\cite{Sigal}. 

Brock~\textit{et al.}~\cite{Brock} recently argued that epigenetic memory might accelerate tumour progression by contributing to the development of drug-resistant cancer cells. In this hypothesis, phenotypic variability from the noisy expression of gene $X$ that confers resistance renders some cells (and their offspring) temporarily insensitive to the drug, thereby increasing the probability of acquiring a mutation conferring permanent immunity. In the present work, we develop a minimal model to study this phenomenon quantitatively. 

To study how gene expression noise impacts the dynamics of isogenic cell populations under stress, we define the reproductive fitness ($W$) as the number
of offspring produced in the presence of the stressor (i.e. a drug) relative to that produced in its absence. For simplicity, we assume that all cells produce offspring at the same rate in the absence of the drug, and define the generation time ($t_D$) as the time it takes for each cell to reproduce once. We set the generation time as unit time and report all time-scales relative to $t_D$. We also assume that cells carry the gene $X$ conferring drug resistance when its expression level $x$ is sufficiently high, and that this gene is expressed stochastically in individual cells.

\begin{figure}[b]
%\begin{center}
\includegraphics*[width=8cm,height=8cm]{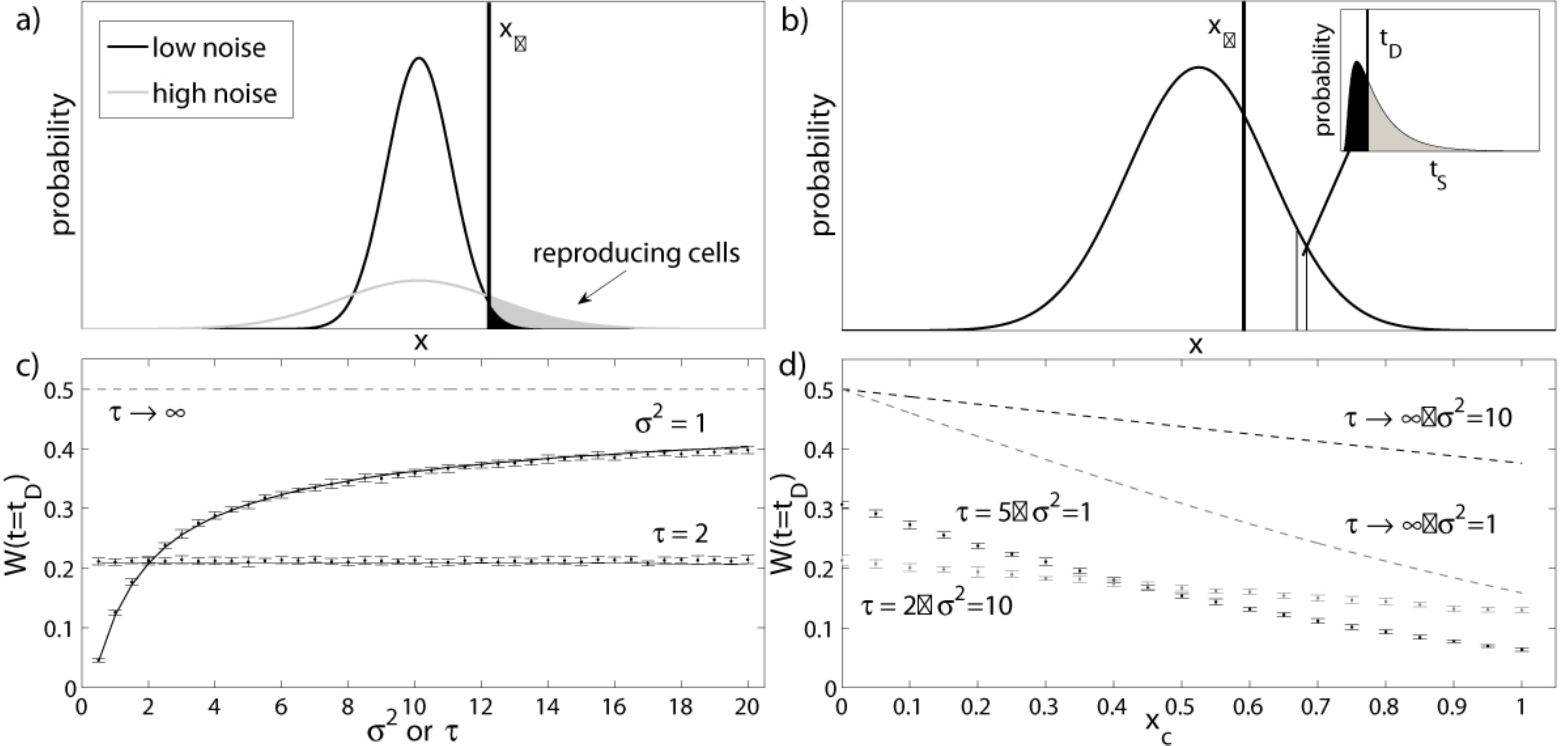}
%\end{center}
\caption{\label{fig1}
Epigenetic effects on a cell population exposed to stress. 
(a) Schematic of instantaneous selection effects. (b) Schematic of generalized model. (c) 
Reproductive fitness at the time of first division $W(t=t_D)$ after the application
of a stress (at $x_c=0$) as a function of $\tau$ or $\sigma^2$ for fixed 
$\sigma^2$ or $\tau$, respectively. Analytical curves (solid lines) were
obtained via numerical solution of Eq.~(\ref{main_eq}). (d) $W(t=t_D)$
as a function of $x_c$ for high and low $\sigma^2$. $\tau$ and $\sigma^2$ are scaled by $t_D$.
Dashed lines represent results obtained from Eq.~(\ref{fitness_eq}), or equivalently Eq.~(\ref{main_eq}) in 
the limit $\tau \rightarrow\infty$.  
}
\end{figure}

The effects of gene expression noise on populations under stress have
previously been analyzed to explain why certain genes have high
expression noise \cite{Fraser,Zhang,Zhuravel}. In these analyses, 
the dependency between gene expression and reproductive fitness was 
defined by the integral
\begin{equation}
W(t)=\int w(x)p_x(x,t)dx,
\label{fitness_eq}
\end{equation}
where $p_x(x,t)$ is the probability distribution function (PDF)
describing the concentration ($x$) of the gene product across the population,
and $w(x)$ is the microscopic fitness function describing the effect
of the drug on the fitness of cells with a given expression level. 
The basic concept is illustrated in Fig.~\ref{fig1}(a) using a model 
where $w(x)$ is described by the Heaviside step function, such that 
cells are unable to reproduce if their expression level is below a critical value, $w(x<x_{c}) = 0$, and
unaffected by the drug otherwise, $w(x \ge x_{c}) = 1$. In this case,
previous theoretical work \cite{Fraser,Zhang,Zhuravel} concluded that high gene expression noise is 
beneficial at high drug-doses, since the fraction of cells expressing above a reproductive threshold $x_c$
increases with the width of the initial expression distribution [Fig.~\ref{fig1}(a)]. 
However, because $p_x(x,t)$ is assumed fixed at 
the time of drug treatment, this conclusion is valid only for
instantaneous selection effects. The analysis of prolonged stress exposure 
necessitates an approach where selection, inheritance and gene
expression dynamics all contribute to the evolution of the population.

%Theory and simulation of epigenetic effects (Figure 1)
%%%%%%%%%%%%%%%%%%%%%%%%%%%%%%%%%%%%%%%%%%%%%%%%%%%%%%%%%%%%%%%%%%%%%%%%%%
%%%%%%%%%%%%%%%%%%%%%%%%%%%%%%%%%%%%%%%%%%%%%%%%%%%%%%%%%%%%%%%%%%%%%%%%%%

Population survival during prolonged drug exposure is a
first-passage time problem. In the absence of mutations conferring
permanent immunity, cells that survive the initial selection will
eventually succumb to the drug since they cannot maintain high
expression indefinitely. Consider a sub-population of cells with the same level of $x$ above $x_c$
[Fig.~\ref{fig1}(b)]. The time interval in which a given cell can reproduce is the
first-passage (or sojourn) time $t_S(x)$, where the threshold $x_c$ represents an absorbing
barrier. Although cells are initially
identical, the expression of the drug-resistance gene evolves
differently in different cells, and the time to reach the
reproductive threshold is a random variable described by the first-passage
time distribution $p_S(x,t_S)$ (Fig.~\ref{fig1}(b), Inset). Since 
only cells with $t_S(x) > t_D$ reproduce, $w(x)$ in Eq.~\ref{fitness_eq} is given by
\begin{equation}
w(x) = \int^{\infty}_{t_{D}}p_S(x,t_S)dt_S',
\label{previous_eq}
\end{equation}
and the overall fitness of the population at time $t$ can be written as
\begin{equation}
W(t) = \int^\infty_{x_c}\left(\int^{\infty}_{t_{D}}p_S(x,t_S)dt_S'
\right)p_x(x,t)dx.
\label{main_eq}
\end{equation}

The population fitness in Equation (\ref{main_eq}) has an explicit solution only in special
cases. Previous analyses~\cite{Fraser,Zhang,Zhuravel} circumvented
this problem, in part, by focusing on initial selection effects ($t
\rightarrow 0$). However, even in this limit, it is also necessary to
assume that all cells above the threshold contribute to fitness (i.e., 
$w(x)=1$ for $x > x_c$). 

To investigate more general cases, we used the 
Ornstein-Uhlenbeck (OU) process to model the level of gene
expression in individual cells \cite{Shahrezaei}. 
This process can be described by the Langevin equation
\begin{equation} \label{ou_eq}
\frac{dx(t)}{dt} = \frac{1}{\tau}\left(\mu-x(t)\right)+c^{1/2}\xi_t,
\end{equation}
where $c$ and $\tau$ are the diffusion constant and the relaxation
time, respectively, and $\xi_t$ is Gaussian white noise ($\left\langle
\xi_t \right\rangle=0$, $\left\langle \xi_t \xi_{t'} \right\rangle=
\delta(t-t')$) \cite{Uhlenbeck}. 
The steady-state PDF of the OU process is a Gaussian distribution with mean $\mu$ and variance $
\sigma^2=c\tau/2$. Without loss of generality, we set $\mu = 0$
and use the fluctuation time-scale $\tau$ to model the time-scale of epigenetic memory.

The fluctuation time-scale of gene expression has been determined experimentally in human lung cancer cells
in terms of the `mixing time' $\tau_m$, defined as  
the lag where the autocorrelation function has decreased by 50\%~\cite{Sigal}.
The mixing time for the stationary OU process is $\tau_m = \tau\ln(2)$. 
The measured values of $\tau_m$ varied  
between~0.5~to~3.0 generations for different genes, corresponding to values of $\tau$ between~0.7~to~4.0~generations for the OU process.

%Effects of epigenetic memory on drug resistance (Figure 2)
%%%%%%%%%%%%%%%%%%%%%%%%%%%%%%%%%%%%%%%%%%%%%%%%%%%%%%%%%%%%%%%%%%%%%%%%%%
\begin{figure}[b]
\begin{center}
\includegraphics*[width=8cm,height=8cm]{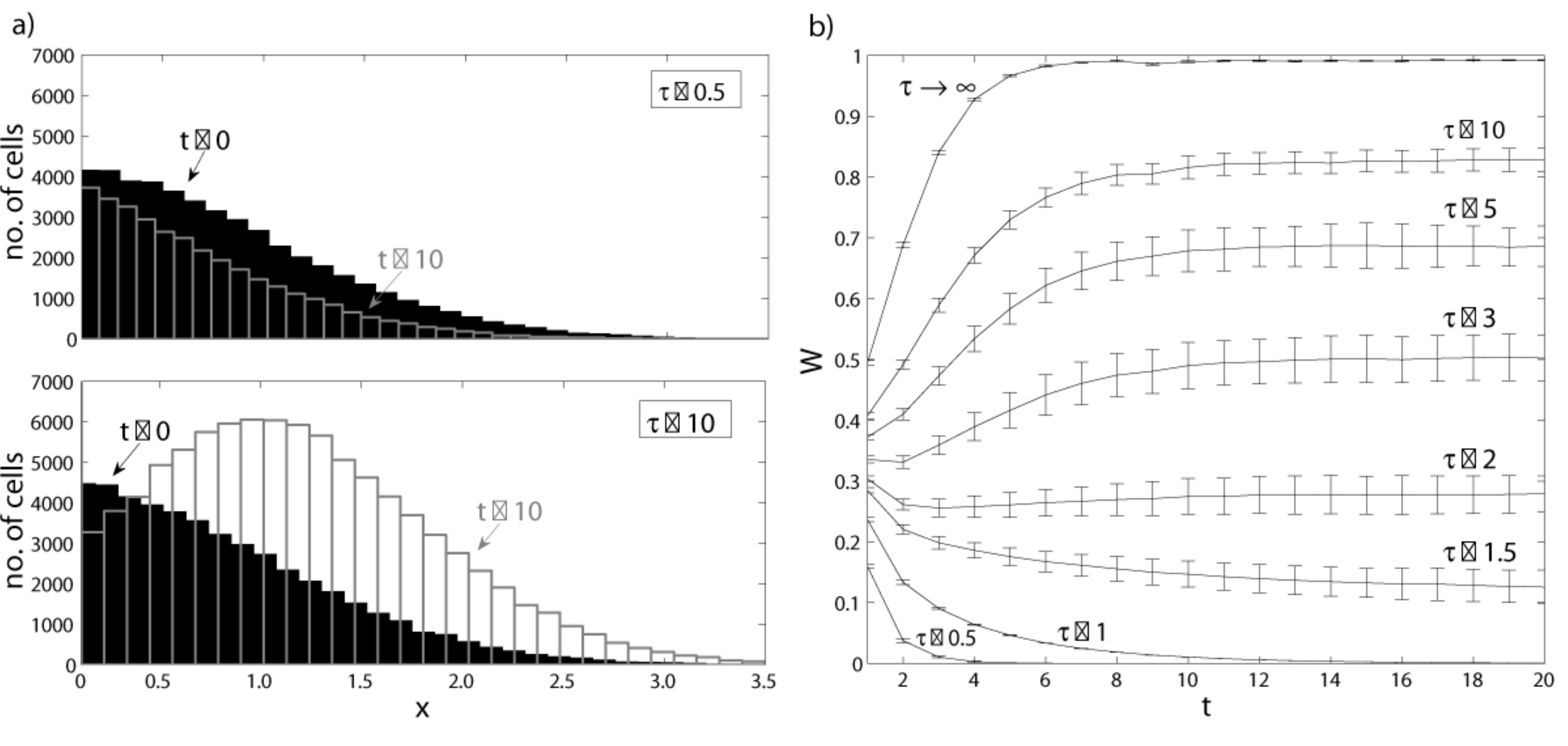}
\end{center}
\caption{\protect
Effect of epigenetic memory $\tau$ on drug resistance at
various timescales. (a) Top and bottom plots show population distributions
corresponding respectively to short ($\tau=0.5$) and long ($\tau=10$) epigenetic memory and
show the fraction of drug resistant cells (i.e., cells with $x > x_c$)
after acute ($t=0$) and a prolonged ($t=10$) drug exposures (single realization of $10^5$ cells). (b) $W$ as a function of
$t$ for various values of $\tau$. $t$, $\tau$ and $\sigma^2$ 
are scaled by $t_D$.  
} \label{fig2}
\end{figure}
%%%%%%%%%%%%%%%%%%%%%%%%%%%%%%%%%%%%%%%%%%%%%%%%%%%%%%%%%%%%%%%%%%%%%%%%%%

First, we examined the effect of drug treatment on
reproductive fitness after one generation time when the
absorbing barrier is located at $x_c=0$. In this case, the first-passage
time PDF for $x>x_c$ is given by~\cite{Wang}
\begin{eqnarray} \label{fpt_eq}
p_S(x,t_S) & = & \frac{x}{\sqrt{2\pi c}} \exp\left(\frac{-x^2 \exp(-
t_S/\tau)}{2c\tau \sinh(t_S/\tau)} + \frac{t_S}{2\tau}\right)
\nonumber \\
& & \times \left(\frac{1}{\tau \sinh(t_S/\tau)}\right)^{3/2}.
\end{eqnarray}
We evaluated the effects of varying the time-scale of epigenetic
memory and the noise amplitude by numerical integration of
Eq.~(\ref{main_eq}), using the steady-state OU distribution to describe
the initial gene expression distribution. Figure~\ref{fig1}(c) shows 
the results for fixed noise ($\sigma^2 = 1$) and variable $\tau
$, and fixed time-scale ($\tau = 2$) and variable $\sigma^2$. 

The time-scale of epigenetic memory significantly affects `acute' reproductive
fitness, even for very long fluctuation relaxation times. For example, when $\tau = 20$, 
$W$ is reduced to~0.4, compared with the value of~0.5 obtained
(irrespectively of the noise amplitude) in the permanent epigenetic memory limit $\tau \rightarrow \infty$ [Fig.~\ref{fig1}(c)].
For $\tau = 2$, the reproductive fitness 
is approximately~0.2, and the majority of cells starting with $x > x_c$ are unable 
to maintain above-threshold gene expression long enough to reproduce.
In this case, the acute reproductive fitness remains constant, presumably 
because changing the noise amplitude for $x_c = 0$ does not change the fraction of cells with $x > x_c$.

To examine cases where $x_c > 0$, it is necessary to use numerical
simulations since a general closed-form solution of the first-passage time PDF
is not available. For this purpose, we employed a population simulation algorithm~\cite{Charlebois2011} in  
which gene expression in each of $N$ individual cells, $x_i(t)$ for
$i=1,\ldots,N$, is obtained by solving Eq.~\ref{ou_eq} numerically~\cite{Gillespie}. 
In these simulations (20 realizations of $10^4$ cells unless indicated otherwise), cell division occurs when a deterministic cell cycle `clock', which is reset at each division, reaches $t_D$. Each cell keeps track of the time since its birth and can only advance its clock if they maintain gene expression above the threshold.
Moreover, cells where $x_i(t) \leq x_c$ are assumed to be fixed and unable to change their
expression level (i.e., $\tau = \infty$). Simulations were initiated
by assigning, to each cell, random initial values of gene expression and
the cell cycle clock from the steady-state distribution of the OU
process and a uniform distribution $[0:t_D]$, respectively. 

Numerical calculations of fitness for $x_c > 0$ identified $\tau$ as a critical determinant of
population survival. Specifically, the fitness of a population with low gene expression noise can 
be greater than that of a population with high noise if the fluctuation relaxation time is sufficiently long.
We observed this in simulations, shown in
Fig.~\ref{fig1}(d), with an increased threshold $x_c$ for
fixed time-scales ($\tau = 2$ or $\tau = 5$) and two different
fluctuation amplitudes ($\sigma^2 = 1$ or $\sigma^2 = 10$). 
When the two populations had the same finite value
of $\tau$, we observed that increased gene expression
noise always provides a fitness benefit (data not shown). 
However, as expected from Fig.~\ref{fig1}(c), 
incorporating stochastic gene expression dynamics (i.e., finite values of $\tau$) generally
yields a significant reduction in fitness compared to the asymptotic permanent memory limit. 
The magnitude of this reduction is sensitive to
both the value of the threshold $x_c$ and the value of $\tau$. This is
illustrated in Fig.~\ref{fig1}(d) where the fitness of the high noise
population is greater than the low noise population only when the
value of $x_c$ is sufficiently high.

In our second case, we analyzed the long-term effects of varying the
time-scale of epigenetic memory on population dynamics and
reproductive fitness. For simplicity, we focus on the case where $x_c = 0$ and noise is fixed 
($\sigma^2 = 1$). Figure~\ref{fig2}(a) shows representative gene
expression distributions obtained after 10 generation times for short-
and long-term epigenetic memory. When the fluctuation time-scale is short ($\tau = 0.5$,  
top panel), the number of cells that may reproduce (i.e., cells with
$x_i(t) > x_c$) is reduced over time since, on average, cells reach
the absorbing barrier faster than they reproduce. Correspondingly,
given enough time, the population will go extinct. This is not the
case when memory is long ($\tau = 10$, bottom panel) and the birth rate
exceeds the rate of loss at the absorbing barrier. In addition, the mode of gene expression distribution shifts to higher values, in
resemblance of experimental observations \cite{Zhuravel}.

Relatively short-term epigenetic memory can result in
permanent drug resistance even in the absence of mutations. This is
illustrated in Fig.~\ref{fig2}(b), which shows how the reproductive
fitness of populations with different memory time-scales evolve over
time. In populations with long-term memory (e.g., $\tau = 5, 10,~
\text{or}~\infty$), the number of cells that may reproduce increases
steady over time and settles in a steady-state where more than half of
them reproduce every generation time (i.e., $W(t) > 0.5$). Importantly, populations
with memory at intermediate time-scales (e.g., $\tau = 1.5, 2,~\text{or}~3$)
may retain long-term viability and finite rates of reproductive fitness.
Because the simulations involve finite populations, the outcome of a  
given realization cannot always be predicted. For example, when $\tau =  
1.5$, a viable population was observed to develop in~29\% of the simulations while
the population went extinct in the remaining~71\% of simulations. While
populations with short memory (e.g., $\tau = 0.5~\text{or}~1$) eventually go extinct,
several cell cycles were needed for the drug to fully effect all cells.

%Effect of epigenetic memory and mutation on cancer populations (Figure 3)
%%%%%%%%%%%%%%%%%%%%%%%%%%%%%%%%%%%%%%%%%%%%%%%%%%%%%%%%%%%%%%%%%%%%%%%%%%
\begin{figure}[b]
\begin{center}
\includegraphics*[width=8cm,height=7cm]{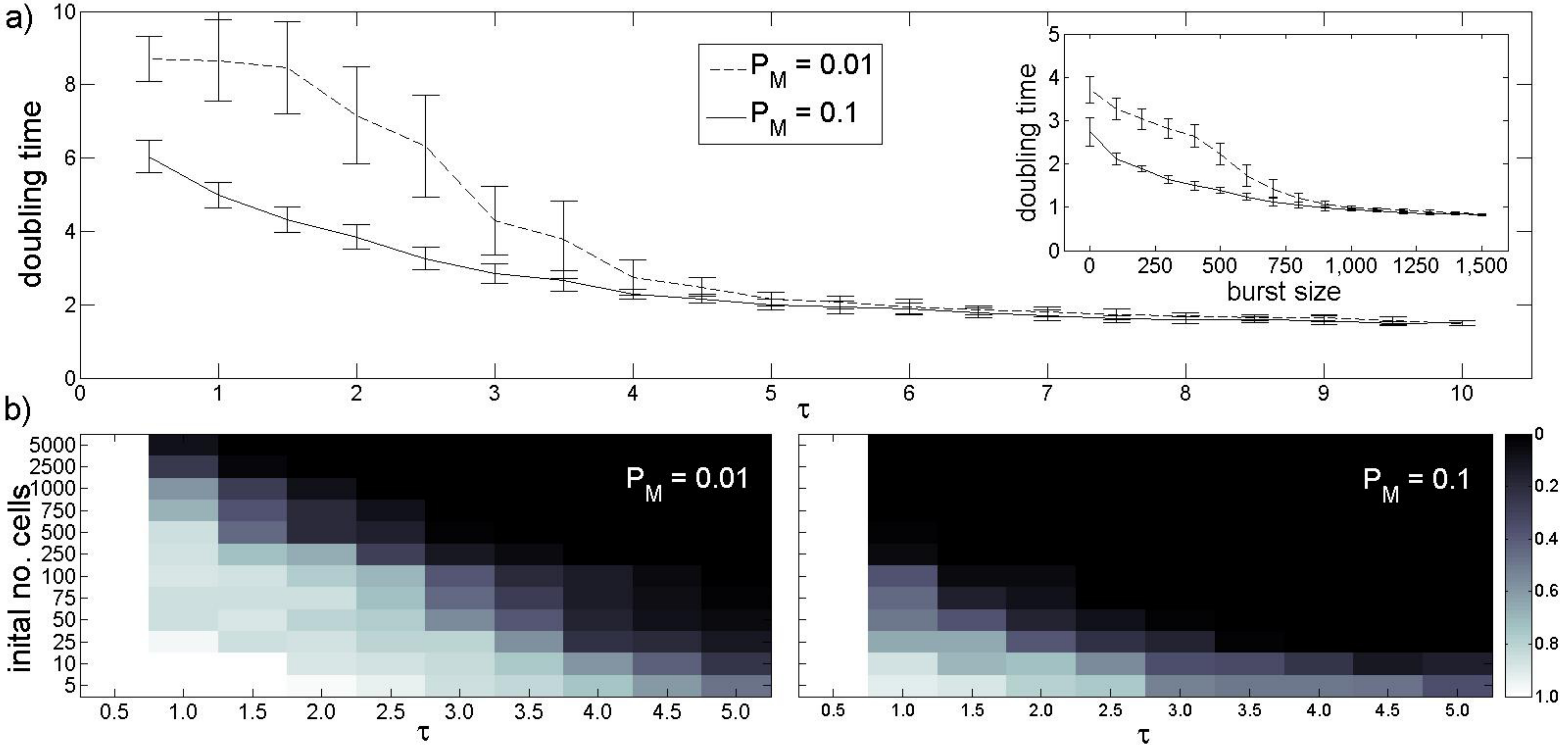}
\end{center}
\caption{\protect
Effect of $\tau$ and $P_M$ on cancer cell populations undergoing
prolonged drug treatment. a) Effect of $P_M$ on doubling time as a function of $\tau$ 
for an initial population of $1000$ non-mutated cells with gene expression
levels above a $95\%$ drug threshold. Simulation results using burst model of gene expression \cite{Jia11} shown in the
inset. b) Heat maps corresponding to (a) show probability of remission after 10 generations (100 realizations). $P_M$, $\tau$ and $\sigma^2$ are scaled by $t_D$. 
} \label{fig3}
\end{figure}
%%%%%%%%%%%%%%%%%%%%%%%%%%%%%%%%%%%%%%%%%%%%%%%%%%%%%%%%%%%%%%%%%%%%%%%%%%

In the third and final case, we investigated the added effect of
genetic mutations on the development of drug-resistance. A central  
element of the Brock \textit{et al.} hypothesis is that temporary
drug resistance due to slow fluctuations in gene expression may contribute
to tumour development by increasing the overall probability that some cells acquire a
mutation conferring permanent immunity. To model this scenario, we allowed each cell with an expression level above
$x_c$ the chance to mutate once per generation time. We denote this probability $P_M$.
If a cell acquired the mutation, it and its offspring were permanently resistant to
the drug, and the survival of a continuously growing population inevitable. 

We first investigated the added effect of mutations on the re-emergence  
of a cancerous tumour under constant drug treatment. In these simulations, we chose $x_c$ such  
that the drug instantaneously removed 95\% of the  
population, and measured the time it took for the remaining cells to  
double in number. Figure~\ref{fig3}(a) shows the dependency of this doubling time  
on $\tau$ when $P_M$ is equal to 0.01 and 0.1. These mutation rates are unrealistically high 
biologically and were chosen to illustrate the effect of epigenetic memory in an extreme limit. 

As expected, increasing the mutation probability significantly reduces the doubling time when  
the gene expression fluctuations are short-lived. Unexpected, however,  
the value of $\tau$ beyond which mutations do not have an additional effect is remarkably  
short despite the unrealistically high mutation rates. Specifically,
the doubling time is more or less unaffected by $P_M$ when $\tau$ is roughly above 4
generations, corresponding to the upper range of mixing times  
observed experimentally~\cite{Sigal}. 

We confirmed our results using a semi-realistic model of gene expression noise~\cite{Jia11} where proteins are synthesized in irregular bursts at irregular intervals (Fig.~\ref{fig3}(a), Inset). We also tested the effect of replacing the fitness threshold with a more realistic sigmoidal fitness function and found no qualitative difference (data not shown). In reality, gene expression dynamics may follow more complex kinetics than that of a simple mean-reverting process due, for example, to multistability and noise-driven switching~\cite{Chang,Kalmar}. Our simulation results demonstrate that such complexity is not required for gene expression noise to have an significant impact on population dynamics under prolonged stress.  

We also determined how the probability of remission depends on the  
mutation rate, the initial number of cancer cells with above-threshold  
expression, and the time-scale of gene expression noise. In these  
simulations, the cancer is in remission if no cells have above  
threshold gene expression and have not acquired a mutation conferring  
permanent immunity within 10 generation times. As expected [Fig.~\ref{fig3}b],  
the probability of remission is greatly decreased when the number of  
initial surviving cancer cells or the mutation rate is increased. Also, when $\tau$
is very short, remission is virtually guaranteed. However, the  
probability that a drug-resistant cell population will emerge can be quite  
substantial within the experimentally observed range of $\tau$. Even  
with a relative low mutation rate ($P_M=0.01$) and 10 surviving cells, the  
probability of remission is only 42\% when $\tau = 4.0$.

In summary, we have analyzed the effect of gene expression noise on the reproductive fitness of isogenic cell  
populations under stress as a first-passage time problem. By explicitly incorporating the  
`epigenetic memory' of this noise (i.e., the fluctuation relaxation time), we have  
generalized previous theoretical work that explained the acute effects of noise amplitude
but did not incorporate gene expression dynamics~\cite{Fraser,Zhang,Zhuravel}. 
This generalization is important for two reasons. First, it has allowed us to demonstrate using a minimal model that 
gene expression noise with biologically realistic time scales has a significant effect on reproductive fitness under stress
and is a critical determinant of population survival. Second, it
enables theoretical and computational investigations of experimentally observed phenomena
associated with prolonged stress exposure, including reversible shifts in gene expression distributions~\cite{Zhuravel}, 
and drug resistance. In this context, we have demonstrated that the time-scale of epigenetic memory  
required to develop a drug-resistant cell population independently of mutations is comparable to that measured  
for certain genes in human cancer cells~\cite{Sigal}. Correspondingly, long-term population survival may not require specialized 
memory-conferring mechanisms. It might, for example, be acheived without a significant fitness cost through bursty gene expression. An important next step is to confirm our findings using more realistic models of gene expression incorporating additional stochastic effects, such as partitioning errors~\cite{Brenner}, and correspondingly, to employ various analytical and numerical methods that may permit solution in these more complex cases (e.g.~\cite{Munsky,Stamatakis}). We anticipate that future analysis of such models will provide a deeper understanding of epigenetic interactions between genes, drugs and population dynamics.

\end{document}